\documentclass[12pt,preprint]{aastex}

\begin{document}

\title{Systematic variation of the residual of PSR B1937+21, the
fingerprint of its companion star?}

\author{Biping Gong}

\affil{ Department of Astronomy, Nanjing University, Nanjing
210093,PR.China}

\email{bpgong@nju.edu.cn}

\begin{abstract}
PSR B1937+21 was the first millisecond pulsar ever measured. Which
has been appeared as a singular pulsar. The high-precision
observation of this pulsar shows systemic long-term variation in
the residual of time of arrivals. This paper modelled the secular
variation by the orbital precession induced time delay of a binary
pulsar system. The fitting requires that as a binary pulsar, PSR
B1937+21 should have small companion star, $m_2\sim
10^{-2}M_{\odot}$, and projected semi-major axis, $x\sim 10^{-4}$s
to $10^{-3}$s. Which corresponds to ignorable radial velocity of
the pulsar to the line of sight. This might explain why it has
been measured as a singular pulsar instead of a binary pulsar.

\end{abstract}
\keywords{
 gravitation---pulsar: individual (PSR B1937+21)---relativity-stars:binaries-stars: fundamental
parameters---stars: neutron}


\section{Introduction}
PSR B1937+21, the first millisecond pulsar and also the fastest
rotating pulsar up to date, was discovered by Backer $\&$ Kulkarni
et al. (1982). Which has been measured as a singular pulsar. The
systemic long-term variation in the residual of time of arrivals
(TOAs) has been measured by different authors, Kaspi et al.
(1994), Lommen $\&$ Backer (2001). The variation may be caused by
planet around the pulsar. However, such possibility was
constrained stringently by Thorsett $\&$ Phillips (1992).

This paper provides  an alternative model to interpret the
systematic variation in TOAs, the orbital precession induced time
delay of a binary pulsar system. Three numerical solutions are
obtained through fitting the residual of TOAs, all of which
demands small companion star mass, $\sim 10^{-2}M_{\odot}$, and in
turn small projected semi-major axis, $x\sim10^{-4}$ or $10^{-3}$.
Therefore, the corresponding radial velocity of the pulsar to the
line of sight (LOS) is so small that it cannot be resolved from
the timing noise of TOAs. Which might explain why the companion
star has not been observed from timing measurement.

This paper shows that although the companion star of PSR B1937+21
cannot be measured directly from the velocity curve, however, the
orbital parameters could still be extracted through the long-time
variation of TOAs.

By the orbital precession model, when the fitted orbital
parameters are put into the standard timing model, the behavior of
PSR B1937+21 as binary pulsar should be similar to PSR J2051-0827
(Doroshenko et al. 2001) and PSR B1957+21 (Arzoumanian et al.
1994), which have similar  companion masses and orbital period.
Moreover, when the additional time delay is also added to the
standard timing model, the systemic variation of TOAs of PSR
B1937+21 should be eliminated, so are PSR J2051-0827 and PSR
B1957+21. PSR B1937+21 might be another millisecond pulsar whose
companion star is evaporated by the radiation of the pulsar as PSR
B1957+21.




\section{The orbital precession model}
Barker $\&$ O'Connell (1975) gives the first gravitational
two-body equation with spins, in which the orbital angular
momentum, vector, ${\bf L}$, is expressed as precesses around a
vector combined by ${\bf S}_1$ and ${\bf S}_2$, and the magnitude
is small 2 Post Newtonian (PN), typically
$1^{\circ}\cdot10^{-4}$yr$^{-1}$, which is insignificant in pulsar
timing measurement.

On the other hand, the precession of orbital angular momentum
vector, ${\bf L}$, around the total angular momentum vector, ${\bf
J}$, in the special case that ignores the spin angular momentum of
one star in a binary pulsar system, has been studied by many
authors (Apostolatos et al. 1994, Kidder 1995, Wex $\&$ Kopeikin
1999). In this case the precession velocity of the orbit plane is
significant, 1.5 PN, typically $1^{\circ}$yr$^{-1}$ for binary
pulsars with orbital period around 10 hours. Which is measurable
in pulsar timing.

Therefore, the two kinds of expressions on the precession velocity
of ${\bf L}$ seems contrary each other. To investigate their
relationship, the first and most important question to be answered
is: which vector should ${\bf L}$ precess around ? The two kinds
of expressions both agree that ${\bf S}_1$ and ${\bf S}_2$, as
well as the vector combined by them precess rapidly (1.5PN) around
the total angular momentum, ${\bf J}$.  Since ${\bf J}$ is at rest
relative to LOS (after counting out the proper motion), so the
small precession velocity of ${\bf L}$ around the vector combined
by ${\bf S}_1$ and ${\bf S}_2$ doesn't mean that precession
velocity of ${\bf L}$ around ${\bf J}$ is also small, and in turn
relative velocity LOS is small.

The situation is analogy to the following case. If the binary
system we observe is replaced by a solar system, one cannot say
that the velocity of planet A is ignorable because its velocity
with respect to planet B is very small. Instead only when the
velocity of planet A is very small relative to the Baryon center
of the solar system which is at rest to the observer is very
small, can one conclude that this velocity is ignorable.

Obviously the role of ${\bf J}$ in a binary system is equivalent
that baryon center in the solar system. And only the velocity
relative to ${\bf J}$ make sense in observation. Therefore, the
precession of ${\bf L}$ should be expressed as around  ${\bf J}$
to compare with observation.

In the one spin case the precession velocity of ${\bf L}$ must be
around ${\bf J}$ and the magnitude must be significant 1.5PN
(Apostolatos et al. 1994, Kidder 1995, Wex $\&$ Kopeikin 1999).
Gong develop the one spin to the general two spins case, and
obtains that still the precession velocity of ${\bf L}$ must be
around ${\bf J}$ and must be significant 1.5PN (Gong1 2003).

The difference is that in the two spins case, the magnitude of the
precession velocity of ${\bf \Omega}_0$ varies rapidly due to the
variation of the sum the spin angular momenta of the two stars,
${\bf S}$, which can lead to significant secular variabilities in
binary pulsars. Whereas, the one spin case predicts a  constant
magnitude of ${\bf \Omega}_0$, and thus it cannot explain the
significant secular variabilities measured in binary pulsars.

Since the spin angular momenta of the two stars in binary pulsar,
$S_1$ and $S_2$, are much smaller than that of the orbital angular
momentum, $L$, therefore, the sum of the two spins, ${\bf S}={\bf
S}_1+{\bf S}_2$, is also much smaller than $L$, or,  $S/L\ll 1$
must be satisfied for a general binary pulsar system. Moreover
since ${\bf J}={\bf L}+{\bf S}$ (${\bf J}$, ${\bf L}$ and ${\bf
S}$ forms a triangle), the constraint, $S/L\ll 1$, means the
misalignment angle between ${\bf J}$ and ${\bf L}$ must be very
small, $\lambda_{LJ}\ll 1$. The conservation of the total angular
momentum of a binary pulsar system, $\dot{\bf J}=0$, leads to
(Barker $\&$ O'Connell 1975),
\begin{equation}
\label{cns} {\bf \Omega}_0 \times {\bf L} = -{\bf \Omega}_1
 \times {\bf S}_1 - {\bf \Omega}_2 \times {\bf S}_2\,,
\end{equation}
the magnitude of the left hand side of Eq($\ref{cns}$) is
$\Omega_{0}L\sin \lambda_{LJ}$. From which one can see that for a
given torque caused by $S_1$ and $S_2$ at the right hand side of
Eq($\ref{cns}$), there are two choices for $\Omega_{0}$. The first
one is that $\Omega_{0}$ is small in the case that
$\lambda_{LJ}\sim 1$, the second one is that $\Omega_{0}$ is large
in the case that $\lambda_{LJ}\ll 1$. As analyzed above, not
matter from the point of view of observation, or from the
constraint that must be satisfied by a binary pulsar ($S/L \ll
1$), $\Omega_{0}$ have to follow the second choice. Which means
${\bf L}$ must precess around ${\bf J}$ with a very small angle of
precession cone, $\lambda_{LJ}\ll 1$, and the magnitude of ${\bf
\Omega}_{0}$ must be significant.


By Eq($\ref{cns}$), the velocity of the orbital precession, ${\bf
\Omega}_0$, which is along ${\bf J}$ and including two spins can
be obtained (Gong1 2003).
\begin{equation}
\label{e1a} \Omega_0= \rho\Omega_{2}\sin\lambda_{LS}+
\rho(\Omega_{1}-\Omega_{2})\frac{S^{\parallel}_{1}}{S}\sin
\lambda^{\parallel}_{LS_{1}}
 \,,
\end{equation}
where $\rho=\frac{1}{\sin\lambda_{JS}}$, and
$S^{\parallel}_{1}=S_1\cos \eta_{SS1}$ denotes the component of
${\bf S}_1$ in the plane determined by $\bf{S}$ and $\bf{J}$.
$\lambda^{\parallel}_{LS_{1}}$ is the angle between
$S^{\parallel}_{1}$ and $\bf{L}$. Note that $L\sin
\lambda_{LJ}\approx \rho S$ is used in Eq($\ref{e1a}$), since
$S/L\ll 1$.  The right-hand side of Eq($\ref{e1a}$) can as well be
written by replacing subscribes 1 with 2 and 2 with 1.  The
precession velocity of the two stars, $\Omega_1$ and $\Omega_2$,
are given by Barker $\&$ O'Connell's equation.

Actually Eq($\ref{e1a}$) can be regarded as transforming the
direction of ${\bf \Omega}_0$ of the Barker $\&$ O'Connell's
equation (two spins) to ${\bf J }$; and it can also be regarded as
add one more spin into the former one spin case given by
Apostolatos et al. (1994), Kidder (1995), Wex $\&$ Kopeikin
(1999).

The scenario of the motion of a binary system has been discussed
by authors, Smarr $\&$ Blandford (1976), Hamilton $\&$ Sarazin
(1982). In which ${\bf L }$, ${\bf S}_1$ and ${\bf S}_2$ all
precess around ${\bf J}$ rapidly (1.5PN). And the relative
velocities of ${\bf L }$ to ${\bf S}_1$ and ${\bf S}_2$ are very
small. From which a slow precession of ${\bf L }$ around ${\bf J}$
is impossible because of the triangle constraint, ${\bf J}={\bf
L}+{\bf S}$. The scenario can solve all the puzzle concerning the
motion of the three vectors, ${\bf L }$, ${\bf S}_1$ and ${\bf
S}_2$.




Since ${\bf S}_1$ and ${\bf S}_2$ precess with different
velocities, $\Omega_{1}$ and $\Omega_{2}$ respectively ($m_1\neq
m_2$), then $\bf{S}$ varies in both magnitude and direction (${\bf
S}_1$, ${\bf S}_2$ and $\bf{S}$ form a triangle), then from the
triangle of $\bf{S}$, $\bf{L}$ and $\bf{J}$,  in react to the
variation of $\bf{S}$, $\bf{L}$ must vary in direction ($|{\bf
L}|=$const), which means the variation of $\lambda_{LJ}$ ($\bf{J}$
is invariable).

The change of $\lambda_{LJ}$ means that the orbital plane tilts
back and forth.  In turn, both $\lambda_{LS}$ and $\lambda_{JS}$
vary with time. Therefore, from Eq($\ref{e1a}$), the derivative of
the rate of orbital precession can be given by,
\begin{equation}
\label{en1} \dot{\Omega}_{0} =\rho\dot{\Omega}+\dot{\rho}\Omega \
\,,
\end{equation}
where $\Omega=\Omega_0/\rho$ and $\dot{\Omega}$ is given by
\begin{equation}
\label{om01} \dot{\Omega}=\Omega_{2}\Omega_{12}X_3X_4-
\Omega_{12}X_1(\Omega_{01}X_2+\Omega_{12}X_3)+\Omega_{12}X_1X_5 \
\,,
\end{equation}
where $\Omega_{12}=\Omega_{1}-\Omega_{2}$,
$\Omega_{01}=\Omega_{1}-\Omega_{0}$,
 $X_1=\frac{S^{\parallel}_{1}}{S}\sin \lambda^{\parallel}_{LS_{1}}$,
 $X_2=\tan\eta_{ss1}$,
 $X_3=\frac{S_{V1}S_{V2}}{S^{2}}\frac{\sin\eta_{s1s2}}{\alpha\sin\lambda_{JS}}
 $,
 $X_4=\frac{\cos^{2}\lambda_{LS}}{\sin \lambda_{LS}}$,
$X_5=\cot\lambda^{\parallel}_{LS_1}\dot{\lambda}^{\parallel}_{LS_1}$
  and
$\dot{\rho}=-\frac{X_3X_4\Omega_{12}}{\sin^{2}\lambda_{JS}}$. In
which $\alpha=\sin \lambda_{JS}+\frac{\cos
^{2}\lambda_{LS}}{\sin\lambda_{LS}}$, $S_{V1}=S_{1}\sin
\lambda_{JS1}$ and $S_{V2}=S_{2}\sin \lambda_{JS2}$ represent
components of $\bf{S}_{1}$ and $\bf{S}_{2}$ that are vertical to
$\bf{J}$.

 Note that $\Omega_1$ and
$\Omega_2$ are unchanged when ignoring the orbital decay (2.5PN),
and $\lambda_{LS_{\alpha}}$ are unchanged also, since they decay
much slower than the orbital decay (Apostolatos et al. 1994).
Through Eq($\ref{en1}$), $\ddot{\Omega}_{0}$ can be easily
obtained (Gong1 2003) .

The S-L coupling in a binary pulsar system results the following
effects. (1) the apsidal motion of the orbital plane which can be
absorbed by $\dot{\omega}$ (Smarr $\&$ Blandford 1976)
\begin{equation}
\label{e3i}
\dot{\omega}^{obs}=\dot{\omega}^{GR}+\Omega_{0}\cos\lambda_{LJ}
\approx\dot{\omega}^{GR}+\Omega_{0} \ \,.
\end{equation}
Notice that $\Omega_{0}$ is a function of time due to the
variation of $S$, as shown in Eq($\ref{e1a}$), whereas,
$\dot{\omega}^{GR}$, the GR prediction of the advance of
periastron is a constant.

(2) the precession of the orbital plane which can be absorbed by
$\dot{x}$,
\begin{equation}
\label{e3c}\dot{x}=-x\Omega_{0}\sin \lambda_{LJ}\sin\eta_0 \cot i
\ \,.
\end{equation}
$|\dot{x}|$ of Eq($\ref{e3c}$) is usually much larger than
$|\dot{x}^{GR}|$, which is caused by the gravitational wave
induced orbital decay.

(3) the nutation of the orbital plane which can be absorbed by
$\dot{P}_b$. The variation in the precession velocity of the orbit
results in a variation of orbital frequency
(${\nu}_{b}=2\pi/P_{b}$),
${\nu}^{\prime}_{b}-{\nu}_{b}=\dot{\Omega}_{0}\Delta t$ . Then
$\dot{\nu}_{b}=\dot{\Omega}_{0}$, which lead to the variation of
$P_b$ (Gong1 2003),
\begin{equation}
\label{ez2} \dot{P}_{b}=-\frac{\dot{\Omega}_{0}P_{b}^{2}}{2\pi} \
\,.
\end{equation}
Notice that the coupling of the spin induced quadrupole moment
with the orbit (Q-L coupling) can also cause apsidal motion which
can explain the secular variabilities on $\dot{\omega}$ and
$\dot{x}$ measured in binary pulsars. However, such effect cannot
explain $\dot{P}_b$, and the second and third order of derivatives
of parameters like, $\ddot{x}$, $\ddot{P}_b$ and ${P}_b^{(3)}$.
Since the Q-L coupling is similar to that of the S-L coupling in
the one spin case, which corresponds to a static precession or
apsidal motion of the orbital plane. In other words, the velocity,
$\Omega_0$ is a constant in these two cases. Whereas, in the two
spins S-L coupling, $\Omega_0$ is a function of time, as given by
Eq($\ref{e1a}$), which can not only explain the parameters that
can be explained by the Q-L coupling or the one spin S-L coupling,
but also explain parameters that cannot be explained by them.

\section{Application to PSR B1937+21}
The essential transformation relating solar system barycentric
time $t_b$ to pulsar proper time $T$ is given by the equation
(Manchester and Taylor 1977)
\begin{equation}
\label{tw1} t_b-t_0=T+\Delta_{R}+\Delta_{E}+\Delta_{S}+\Delta_{A}
 \,,
\end{equation}
where $\Delta_{R}$ is the "Roemer time delay", which corresponds
to the propagation time across the binary orbit; $\Delta_{E}$ and
$\Delta_{S}$ are the orbital Einstein and Shapiro delays; and
$\Delta_{A}$ is a time delay related with aberration caused by
rotation of the pulsar. The dominant term concerning the orbital
effects is the Roemer delay, $\Delta_{R}$, given by (Taylor $\&$
Weisberg 1989)
\begin{equation}
\label{tw2} \Delta_{R}=xF(\omega+u)
 \,,
\end{equation}
where
\begin{equation}
\label{tw2a} F(\omega+u)=\sin \omega [\cos u-e(1+\delta_r)]+
[1-e^{2}(1-\delta_{\theta})^{2}]^{1/2}\cos \omega\sin u
 \,,
\end{equation}
where $u$ is the eccentric anomaly, and $\omega$ is the angular
distance periastron from the node. In calculation, the small
quantities $\delta_{r}$ and $\delta_{\theta}$ due to aberration
are ignored.

When the contribution of orbital precession to the time of arrival
is considered, the dominant term that absorbs the additional time
delay is also the Roemer delay. Which can deviate from the
standard one,
 $\Delta_{R}$, and therefore, leads to an additional time delay,
\begin{equation}
\label{opt2}\Delta = xF(\omega^{\prime}+u^{\prime})- xF(\omega+u)
=2x\sin(\Omega_0t/2)\cos\theta\approx x\Omega_0t\cos\theta
 \,.
\end{equation}
where $ \theta=(\dot{\omega}+\Omega_0)t/2+\omega_i$. The
additional time delay, $\Delta$, given by  Eq($\ref{opt2}$), is  a
function of time due to the variation of $\Omega_0$. Which can
contribute the TOAs via Eq($\ref{tw1}$). And such dynamic effect
might be interpreted as effect of propagation, like  $\dot{DM}$ in
some binary pulsars. The effect of $\dot{\Omega}_0$,
$\ddot{\Omega}_0$ can be absorbed by such parameters as
$\dot{P}_b$, $\ddot{P}_b$, $\ddot{x}$, ${x}^{(3)}$, $\ddot{DM}$,
${DM}^{(3)}$, etc (Gong2, Gong3, 2003). The orbital precession
model predicts that
$\frac{{P}_b^{(3)}}{\ddot{P}_b}\sim\frac{\ddot{P}_b}{\dot{P}_b}\sim\Omega_2
$ for a same binary pulsar, so is the derivatives of $x$ and $DM$.
Evidences of such correlation can be found in binary pulsars, such
as PSR J2051-0827 (Doroshenko et al. 2001) and PSR B1957+21
(Arzoumanian et al. 1994), PSR B1534+12 (Konacki et al. 2003).
Further more the orbital precession model predicts that binary
pulsars with long orbital period, i.e., for days, correspond to
much smaller $\Omega_0$ and $\Omega_2$, and in turn much smaller
derivatives of $x$, $P_b$ and $DM$ by Eq($\ref{e3c}$),
Eq($\ref{ez2}$) and Eq($\ref{opt2}$). Which make the corresponding
derivatives of such binary pulsar more difficult to observe than
that of binary pulsars with small $P_b$, i.e., a few hours. Such
correlation can also be find in binary pulsars

As shown in Fig~1, the amplitude of the measured time delay is
about $10^{-6}$s. By simple estimation, if
$x\Omega_0\cos\theta\approx 10^{-13}$ in Eq($\ref{opt2}$), then
the orbital precession induced time delay can be $\Delta\approx
10^{-6}$s also. There are two possibilities to satisfy this
relation: (1) small $\Omega_0$ and $\dot{\omega}$ like PSR
B1855+09 (large $P_b$), and large $x$, however in such case, it
will take much longer time to finish a variation like that of
Fig~1. And (2) large $\Omega_0$ and $\dot{\omega}$, i.e., $\sim
10^{\circ}$yr$^{-1}$, and small $x$, obviously case (2) can not
only satisfy the amplitude but also the periodicity of variation.
Therefore, the orbital precession model implies that PSR B1937+21
should have small $x$, and small $P_b$ (which corresponds to large
$\Omega_0$ and $\dot{\omega}$).

The vectors ${\bf S}_1$, ${\bf S}_2$ and ${\bf S}$ are studied in
the coordinate system of the total angular momentum, in which the
z-axis directs to ${\bf J}$, and the x- and y-axes are in the
invariance plane. ${\bf S}$ can be represented by $S_{P}$ and
$S_{V}$, the components parallel and vertical to the z-axis,
respectively:
\begin{equation}
\label{ex1}
 S=(S_{V}+S_{P})^{1/2} \ .
\end{equation}
$S_{P}$ and $S_{V}$ can be expressed (recall $S_{V1}$, $S_{V2}$
and $S_{V}$ form a triangle) as
$$S_{P}=S_{1}\cos \lambda_{JS1}+S_{2}\cos \lambda_{JS2} \ ,$$
\begin{equation}
\label{ex2}  S_{V}= (S_{V1}^{2}+S_{V2}^{2}-2S_{V1}S_{V2}\cos
\eta_{S1S2})^{1/2} \ ,
\end{equation}
where $\eta_{S1S2}$ is the misalignment angle between $S_{V1}$ and
$S_{V2}$, which can be written as
\begin{equation}
\label{ex5} \eta_{S1S2}=(\Omega_1-\Omega_2)t+\phi_{i}  \ .
\end{equation}
Therefore, by the variation of $S$ as function of time (in the
case of one spin, $S=$const), one can obtain $\Omega_0$ as
function of time through Eq($\ref{e1a}$). Thus the measured
$\Delta^{obs}$ of Fig~1 can be fitted step by step via  the
orbital precession induced  $\Delta$ given by Eq($\ref{opt2}$),
\begin{equation}
\label{ex5a}
\Delta^{obs}(t_{k+1})-\Delta^{obs}(t_k)=\Delta(t_{k+1})-\Delta(t_k)
\ .
\end{equation}
Note that both $\Omega_0$ and $\cos\theta$ are functions of time
in Eq($\ref{opt2}$). Which are responsible for variation of
residual in Fig~1.

Three numerical solutions can be obtained by the Monte Carlo
solution of Eq($\ref{ex5a}$). The solutions are shown in Table
1,3,5, respectively.  The three solutions shows that $P_b$ is from
$0.0708$d to $0.10$d, and $x$ from $3.73\cdot 10^{-4}$s to
$4.25\cdot 10^{-3}$s, and $m_2$ from $0.0126M_{\odot}$ to
$0.0373M_{\odot}$. The possibility of other solutions cannot be
excluded.


The standard DD timing model uses 12 parameter concerning the
orbit to fit the delays of Eq($\ref{tw1}$) (Taylor $\&$ Weisberg
1989). The total number of binary parameters of the orbital
precession model as shown in Table 1 is also 12. There are some
parameters in common, such as, $e$, $P_b$, $\omega_0$ and $T_0$.
The difference is that the latter can cause an additional time
delay, as shown in Eq($\ref{opt2}$), through the parameters of the
second row of Table 1 (including $m_1$ and $m_2$ of the first
row). Whereas, the former doesn't have this delay. In other words,
the latter only adds one more delay term to the right hand side of
Eq($\ref{tw1}$), and all other terms (effects) are the same as the
former models. The relation of the orbital precession model with
the DDGR model can be given by
 $\dot{P}_b^{obs}=\dot{P}_b^{GR}+\dot{P}_b^{OP}$, and by
 Eq($\ref{e3i}$).

By the results of Table 1,3,5 and the orbital precession model,
one can easily obtain the following Post-Keplerian parameters.
Notice that Table 2,4,6 correspond to the solution 1,2,3,
respectively.

\section{discussion}
The orbital precession model predicts that PSR B1937+21 is a
binary pulsar with very small orbital period, $P_b=1.7$hr. Then
there is a question, why it has been measured as a singular
pulsar? Actually it can be answered by the fitted orbital
parameters of Table 1. The radial velocity (to line of sight) of
the pulsar is given (Shapiro $\&$ Teukolsky 1983),
\begin{equation}
\label{conc1} {\bf v}_1\cdot{\bf n}=\frac{2\pi a_1\sin
i}{P_b(1-e^2)^{1/2}}[\cos(\omega+u)+e\cos\omega] \ ,
\end{equation}
where $a_1=\frac{m_2a}{m_1+m_2}$. Then the observed pulsar period
deviates from the intrinsic period by a small amount,
\begin{equation} \label{conc2} P^{obs}=P^{int}(1+\frac{{\bf
v}_1\cdot{\bf n}}{c}) \approx P^{int}(1+\frac{2\pi x}{P_b})\ .
\end{equation}
Therefore, the contribution of the orbital precession effect to
$P^{obs}$  is about $\delta P^{OP}\approx2\pi {x}P/{P_b}\approx
1.6\cdot10^{-20}$s (corresponds to solution 1). Whereas, the last
decimal of pulsar period, $P$, is $4(2)\cdot10^{-18}$s (Kaspi et
al. 1994), which is approximately two order of magnitude larger
than $\delta P^{OP}$. In other words, the modulation of the radial
velocity of the pulsar (induced by binary motion) to the TOAs is
ignorable. Moreover the small orbital inclination angle
($i\sim10^{-2}$) given by solution 1 and 2, means that the orbital
plane of the binary may be face on, then it is also impossible to
observe the eclipse of this binary pulsar. These factor may make
it difficult to observe the effects of the companion star.

By the orbital precession model, a binary pulsar with very small
orbital period, i.e., a few hours, can produce a long-term time
delay i.e., years, as given by Eq($\ref{opt2}$). Therefore, the
companion star of PSR B1937+21, although unmeasurable in the
orbital period time scale, due to small $x$, might imprint in the
long-term measurement of TOAs.

The results of the three numerical solutions can be tested by
TEMPO. Put the orbital parameters, such as $P_b$, $x$, $e$ of
Table 1 (3,5), as well as Post-Kepler parameters $\dot{P}_b$,
$\dot{\omega}$ of Table 2 (4,6) into the corresponding terms in DD
model.  Then  behavior of PSR B1937+21 should be very similar to
PSR B1957+20 and PSR J2051-0827, since they have similar companion
mass and close orbital period. Moreover, when the orbital
precession induced  time delay of Eq($\ref{opt2}$) is added into
Eq($\ref{tw1}$), then the systematic variation of residuals of
TOAs should be eliminated. Since different initial time
corresponds to different phase angles, so the initial phases, like
$\omega_i$, $\phi_i$ may need to be adjusted when fitted by TEMPO.

Assuming PSR J2051-0827 and PSR B1957+20 are fitted as a singular
pulsar, then they should have similar the residual of TOAs (after
counting out DM variation) as that of PSR B1937+21. And  binary
pulsars such as, PSR J0437-4715, with much larger $P_b$, should
take much longer time to finish the similar variation in residuals
of TOAs like PSR B1937+21, in the case that PSR J0437-4715 is
fitted as a singular pulsar.

The evolutionary links between millisecond pulsars and their
binary progenitor systems involve a number of exotic astrophysical
phenomena. In their late stages of evolution, neutron stars in
low-mass X-ray and pulsar binaries may evaporate their companions
through the strength of their radiation (Alpar et al. 1982,
 Bhattacharya $\&$ van den Heuvel
1992, King et al 2003). PSR B1957+20 has provided strong evidences
to that scenario,  the presumed binary pulsar PSR B1937+21 might
provide another one.

By the spin angular momentum of the pulsar given in Table 1,3,5,
one can easily obtain its moment of inertia, $I_1=1.5I_{45}$,
$2.6I_{45}$, $0.44I_{45}$ for solution 1,2,3, respectively. (since
pulsar period is $P=1.5578$ms), which are consistent with the
prediction of neutron star equation of state. It is interesting
that PSR B1937+21 having the smallest pulsar period, $P$, might
also has the smallest  $x$ of all millisecond pulsars.

\acknowledgments I would like to thank Prof. T.Lu for constructive
suggestions. I also thank Prof.Z.R.Wang, X.D.Li, Z.Y.Li, Z.D.Dai,
D.M.Wei for helpful discussion on the manuscript.

\clearpage
\begin{table}
\begin{center}
\caption{Solution 1, obtained by fitting the measured data of
Fig~1}
\begin{tabular}{lllllllll}
\hline \hline   $m_1(M_{\odot})$ & $m_2(M_{\odot})$ &
 $e$ &  $P_b$(d) & $i$ & $T_0$(yr)\\
$1.30$ & $0.0373$ &
 $0.0426$ & $0.0708$ & $7.25\cdot10^{-3}$  & $0.779$ \\\hline

$S_1$(g~cm$^{2}$s$^{-1}$) & $S_2$(g~cm$^{2}$s$^{-1}$) &
$\lambda_{JS_{1}}$ & $\lambda_{JS_{2}}$ & $\phi_i$ & $\omega_{i}$

 \\
 $1.80\cdot10^{48}$ & $1.83\cdot 10^{48}$
& $0.496$ & $0.0677$ & $0.983$ & $2.56$
\\ \hline\hline
\end{tabular}
\end{center}
{\small
All angles are in radian. $S_1$ corresponds to the moment of
inertia of the pulsar, $I_1=1.53I_{45}$.
 }
\end{table}
\begin{table}
\begin{center}
\caption{Derived parameters through solution 1}
\begin{tabular}{lllll}
\hline \hline   $\Omega_0$(s$^{-1}$) & $x$(s) & $\dot{x}$ &  $\ddot{x}$(s$^{-1}$) & $\dot{x}^{GR}$\\
$-5.16\cdot10^{-10}$ & $3.73\cdot10^{-4}$ & $8.33\cdot10^{-14}$ &
$-4.37\cdot10^{-22}$ & $3.28\cdot10^{-21}$
\\\hline

$\dot{\omega}^{GR}$(s$^{-1}$) & $\dot{P}_b$ &
$\ddot{P}_b$(s$^{-1}$)  & $P_b^{(3)}$(s$^{-2}$) & $\dot{P}_b^{GR}$\\

 $1.11\cdot10^{-8}$ & $1.43\cdot 10^{-11}$ & $2.52\cdot10^{-18}$
 & $-2.96\cdot 10^{-27}$  & $8.06\cdot 10^{-14}$\\ \hline\hline
\end{tabular}
\end{center}
{\small $\dot{x}$ and $\ddot{x}$ are calculated by assuming
$\eta_0=\omega_0$
 }
\end{table}
\begin{table}
\begin{center}
\caption{Solution 2}
\begin{tabular}{lllllllll}
\hline \hline   $m_1(M_{\odot})$ & $m_2(M_{\odot})$ &
 $e$ &  $P_b$(d) & $i$ & $T_0$(yr)\\
$1.38$ & $0.0364$ &
 $0.147$ & $0.100$ & $0.0101$  & $0.779$ \\\hline

$S_1$(g~cm$^{2}$s$^{-1}$) & $S_2$(g~cm$^{2}$s$^{-1}$) &
$\lambda_{JS_{1}}$ & $\lambda_{JS_{2}}$ & $\phi_i$ & $\omega_{i}$

 \\
 $3.04\cdot10^{48}$ & $1.44\cdot 10^{48}$
& $1.46$ & $-0.602$ & $3.91$ & $4.15$
\\ \hline\hline
\end{tabular}
\end{center}
{\small
All angles are in radian. $S_1$ corresponds to the moment of
inertia of the pulsar, $I_1=2.59I_{45}$.
 }
\end{table}
\begin{table}
\begin{center}
\caption{Derived parameters through solution 2}
\begin{tabular}{lllll}
\hline \hline   $\Omega_0$(s$^{-1}$) & $x$(s) & $\dot{x}$ &
$\ddot{x}$(s$^{-1}$) & $\dot{x}^{GR}$\\

$-1.94\cdot10^{-10}$ & $6.16\cdot10^{-4}$ & $-6.63\cdot10^{-14}$ &
$9.11\cdot10^{-22}$ & $2.49\cdot10^{-21}$
\\\hline

$\dot{\omega}^{GR}$(s$^{-1}$) & $\dot{P}_b$ &
$\ddot{P}_b$(s$^{-1}$)  & $P_b^{(3)}$(s$^{-2}$) & $\dot{P}_b^{GR}$\\

 $6.59\cdot10^{-9}$ & $3.19\cdot 10^{-11}$ & $-3.07\cdot10^{-20}$
 & $2.14\cdot 10^{-28}$  & $5.22\cdot 10^{-14}$\\ \hline\hline
\end{tabular}
\end{center}
{\small $\dot{x}$ and $\ddot{x}$ are calculated by assuming
$\eta_0=\omega_0$
 }
\end{table}
\begin{table}
\begin{center}
\caption{Solution 3}
\begin{tabular}{lllllllll}
\hline \hline   $m_1(M_{\odot})$ & $m_2(M_{\odot})$ &
 $e$ &  $P_b$(d) & $i$ & $T_0$(yr)\\
$1.88$ & $0.0126$ &
 $0.414$ & $0.0975$ & $0.250$  & $0.779$ \\\hline

$S_1$(g~cm$^{2}$s$^{-1}$) & $S_2$(g~cm$^{2}$s$^{-1}$) &
$\lambda_{JS_{1}}$ & $\lambda_{JS_{2}}$ & $\phi_i$ & $\omega_{i}$

 \\
 $5.15\cdot10^{47}$ & $1.85\cdot 10^{47}$
& $1.34$ & $0.0474$ & $3.34$ & $0.687$
\\ \hline\hline
\end{tabular}
\end{center}
{\small
All angles are in radian. $S_1$ corresponds to the moment of
inertia of the pulsar, $I_1=0.439I_{45}$.
 }
\end{table}
\begin{table}
\begin{center}
\caption{Derived parameters through solution 3}
\begin{tabular}{lllll}
\hline \hline   $\Omega_0$(s$^{-1}$) & $x$(s) & $\dot{x}$ &
$\ddot{x}$(s$^{-1}$) & $\dot{x}^{GR}$\\

$2.95\cdot10^{-11}$ & $4.25\cdot10^{-3}$ & $-9.90\cdot10^{-16}$ &
$1.44\cdot10^{-23}$ & $2.02\cdot10^{-20}$
\\\hline

$\dot{\omega}^{GR}$(s$^{-1}$) & $\dot{P}_b$ &
$\ddot{P}_b$(s$^{-1}$)  & $P_b^{(3)}$(s$^{-2}$) & $\dot{P}_b^{GR}$\\

 $9.86\cdot10^{-9}$ & $-4.84\cdot 10^{-12}$ & $3.39\cdot10^{-20}$
 & $-5.62\cdot 10^{-29}$  & $5.99\cdot 10^{-14}$\\ \hline\hline
\end{tabular}
\end{center}
{\small $\dot{x}$ and $\ddot{x}$ are calculated by assuming
$\eta_0=\omega_0$
 }
\end{table}

\begin{figure}[t]

\begin{center}
\includegraphics[87,87][700,700]{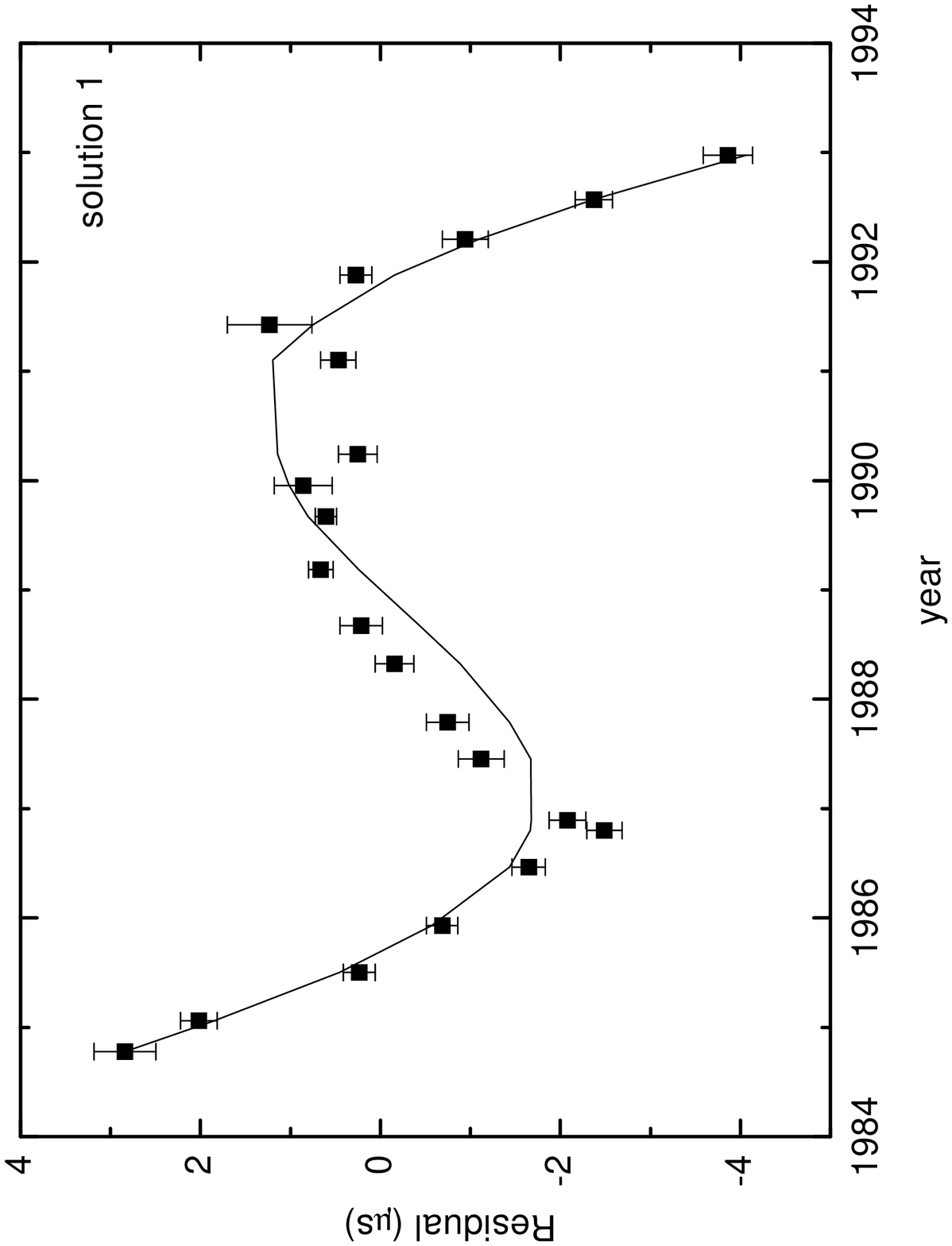}

\end{center}

\caption{The scattered points are measured by Kaspi et al. (1994),
the solid line is given by the orbital precession model.}
\end{figure}

\begin{figure}[t]

\begin{center}
\includegraphics[87,87][700,700]{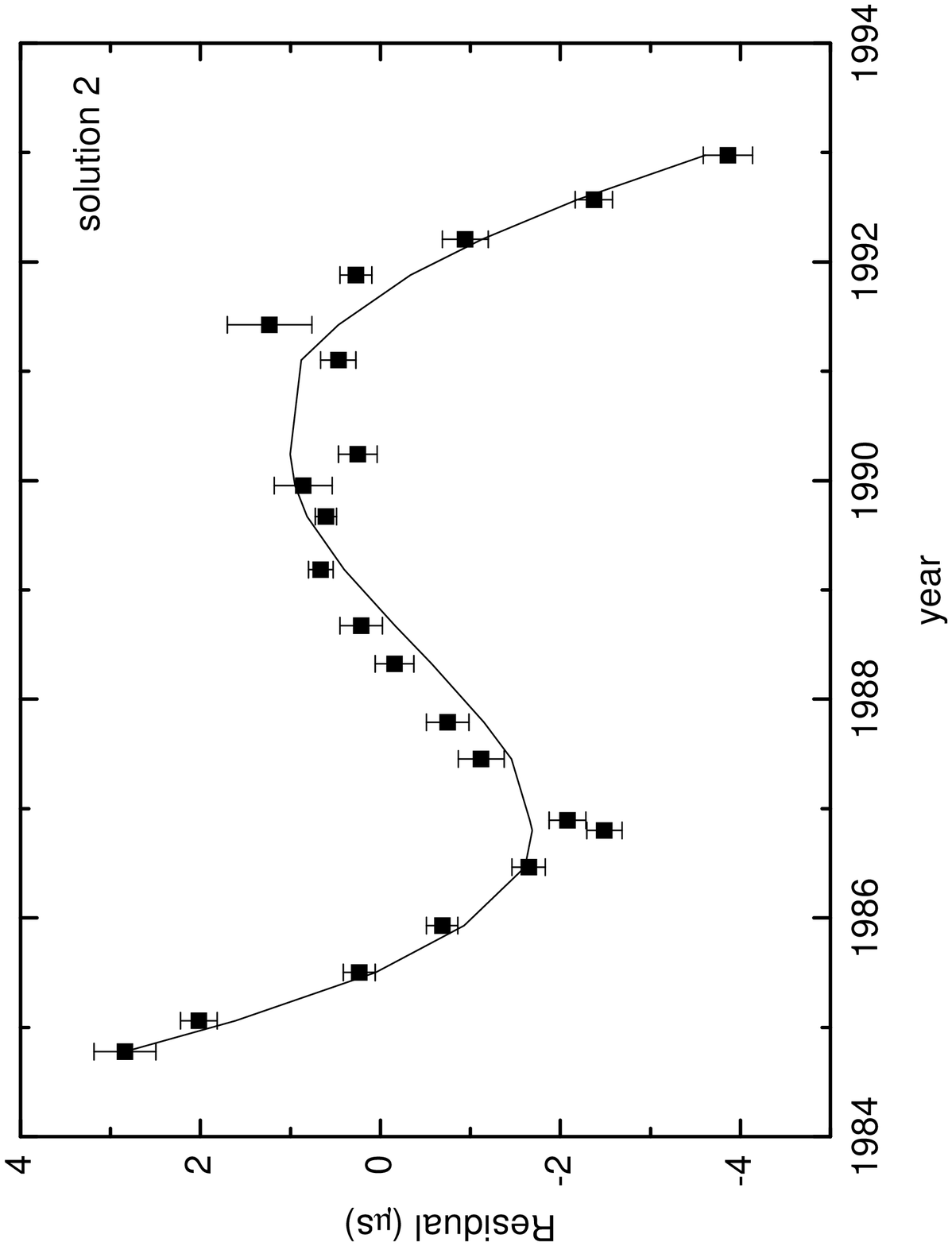}

\end{center}

\caption{The scattered points are measured by Kaspi et al. (1994),
the solid line is given by the orbital precession model.}
\end{figure}
\begin{figure}[t]

\begin{center}
\includegraphics[87,87][700,700]{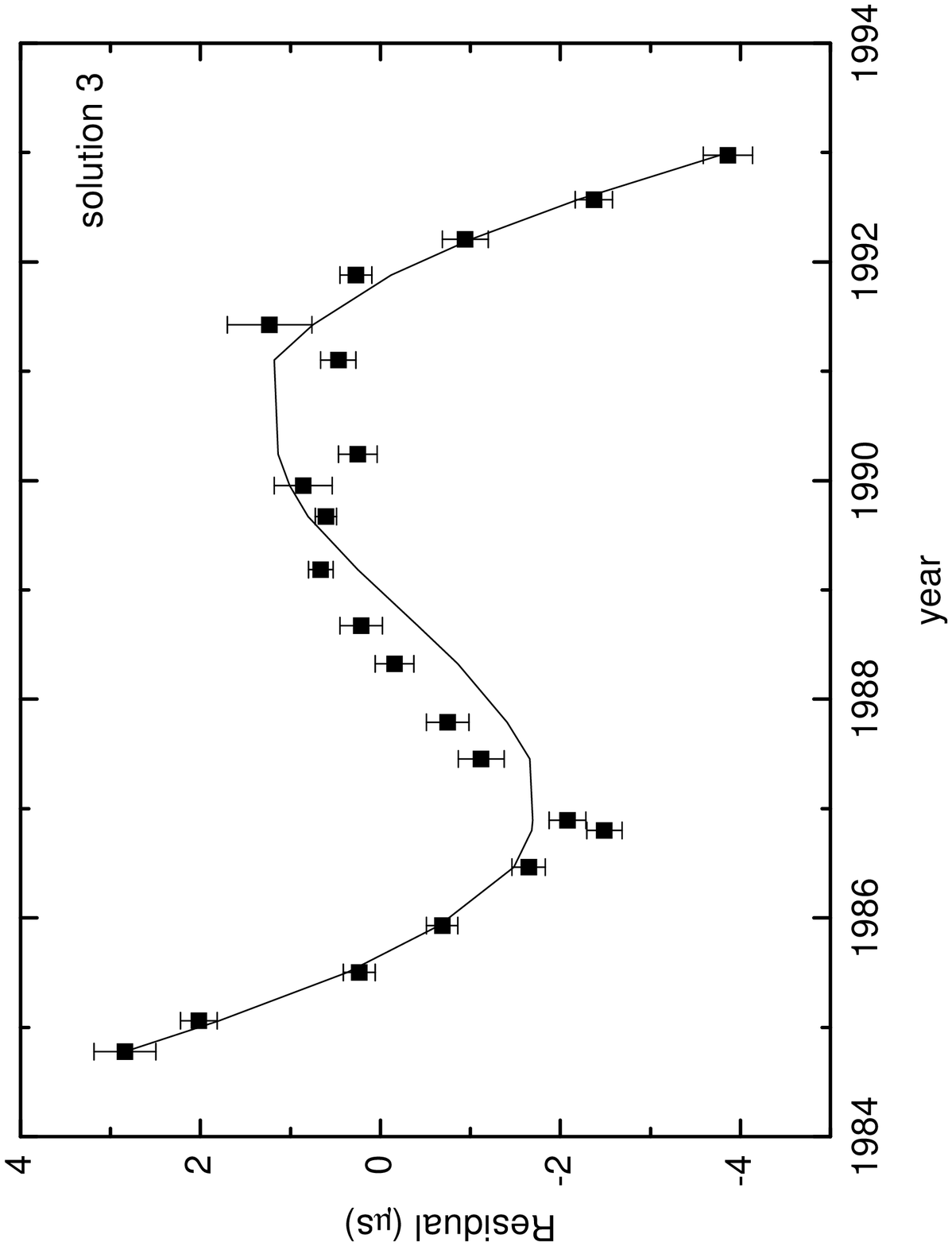}

\end{center}

\caption{The scattered points are measured by Kaspi et al. (1994),
the solid line is given by the orbital precession model.}
\end{figure}

\end{document}